\documentclass[prb,twocolumn,floatfix]{revtex4-1}
\usepackage{amsmath,amscd}
\usepackage{bm}
\usepackage{graphicx}
\usepackage{amssymb}
\usepackage{datetime}
\usepackage{color}
\usepackage{sidecap}
\usepackage{braket}
\usepackage{times}
\newcommand{\vb}[1]{\boldsymbol{#1}}
\begin{document}
\newcommand{\figdir}{.}
\newcommand{\figwidth}{0.9\columnwidth}
\newcommand{\ffigwidth}{0.4\columnwidth}
%
\title{Multifractal finite-size scaling at the Anderson transition in the unitary symmetry class}
%
\newcommand{\freiburg}{Physikalisches Institut, Albert-Ludwigs-Universit\"{a}t Freiburg, Hermann-Herder-Stra{\ss}e 3, D-79104, Freiburg, Germany}
\author{Jakob Lindinger}
\author{Alberto Rodr\'iguez}
\email[]{Alberto.Rodriguez.Gonzalez@physik.uni-freiburg.de}
\affiliation{\freiburg}

\begin{abstract}
We use multifractal finite-size scaling to perform a high-precision numerical study of the critical properties of the Anderson localization-delocalization transition in the unitary symmetry class, considering the Anderson model including a random magnetic flux. We demonstrate the scale invariance of the distribution of wavefunction intensities at the critical point and study its behavior across the transition.    
Our analysis, involving more than $4\times10^6$ independently generated wavefunctions of system sizes up to $L^3=150^3$, yields accurate estimates for 
the critical exponent of the localization length, $\nu=1.446 (1.440,1.452)$, the critical value of the disorder strength and the multifractal exponents. 
\end{abstract}
\maketitle

\section{Introduction}
\label{sec:introduction}
At the critical point of the Anderson transition (AT), the single-particle eigenstates take an exotic form: 
Each spatial iso-`surface'
of the wavefunction intensity within the system is a fractal with a certain fractal dimension.\cite{Aoki1983,Aoki1986,Evers2008} 
The composition of all these fractals comprises a multifractal wavefunction whose 
features are part of the fingerprint of the critical point, and thus they are shared by all models belonging to the same universality class. 
Most interestingly, the study of multifractality in the wavefunctions across the critical region provides an efficient numerical tool ---multifractal finite-size scaling (MFSS)--- to monitor and quantitatively characterize the transition.\cite{Rodriguez2010,Rodriguez2011} This technique has provided insight into the localization-delocalization transition in different models and  symmetry classes,\cite{Huang2013,Ujfalusi2015a} into the quantum percolation problem, \cite{Ujfalusi2014} and it has also revealed the
existence of multifractality in the spectrum of the Dirac operator in quantum chromodynamics.\cite{Ujfalusi2015b}

The MFSS formalism bears the potential to expose the effect of many-particle interactions on the critical properties of the AT,\cite{Richardella2010,Burmistrov2013b,Amini2014,Burmistrov2015a} and work along this line is currently being pursued. \cite{Harashima2012,Harashima2014,Carnio2017} 
Indeed, 
the significance of multifractality goes well beyond non-interacting models: It has been found that the ground state wavefunction of certain spin systems exhibits multifractality in Hilbert space in the absence of any disorder.\cite{Stephan2009,Stephan2010,Stephan2011,Atas2012,Atas2013} 
It is furthermore possible to distinguish between different quantum spin phases in these many-body systems by studying corrections to multifractal scaling.\cite{Luitz2014,Misguich2017}
Multifractality also seems to play a prominent role in interacting systems subject to strong disorder,\cite{DeLuca2013,Luitz2015} in which an insulating phase can emerge, corresponding to a many-particle wavefunction which is localized in Fock space.\cite{Basko2006c,Basko2006b,Gornyi2005}
This has revived the interest for the AT in complex geometries and random graphs,\cite{DeLuca2014,Kravtsov2015,Tarquini2016} leading to a controversy about the existence of a non-zero measure phase populated by multifractal (delocalized non-ergodic) states. \cite{Altshuler2016,Altshuler2016a,Tikhonov2016,Tikhonov2016a,Garcia-Mata2016} 
The study of multifractality and of the applicability of MFSS to different models is therefore of primary importance not only for disordered systems, but also for the understanding of quantum many-body systems. 

Here, we present a high-precision numerical analysis of the AT in the unitary symmetry class, one of the ten existing symmetry classes for disordered systems.\cite{Evers2008} The defining feature of the unitary symmetry class is the absence of time reversal symmetry, which can be broken by applying an external magnetic field or by the presence of magnetic impurities. The question of how the AT is affected by a certain concentration of magnetic moments is currently under investigation and it is potentially relevant to understand the experimental observations of metal-insulator transitions in doped semiconductors.\cite{Jung2016,Kettemann2012,Kettemann2009}
The MFSS formalism has already been applied in the unitary symmetry class by Ujfalusi and Varga in Ref.~\onlinecite{Ujfalusi2015a}. In this work we consider a different Hamiltonian in order to (i) confirm the values of multifractal and critical exponents in this universality class, (ii) present the study of the behavior of the scaling of the probability density function (PDF) of wavefunction intensities (which is currently lacking in the $d=3$ unitary symmetry class) and demonstrate its scale invariance at the critical point in the absence of time-reversal symmetry, and (iii) increase the precision and reliability of the analysis by considering larger system sizes up to $L=150$ and more statistics. 

The organization of the paper is as follows: In Section \ref{sec:model}, we describe briefly the Hamiltonian considered, which combines the effects of a scalar disordered potential and a random magnetic flux. In Section \ref{sec:mf}, we recall the basics of multifractality at the critical point of the AT. In Section \ref{sec:mffluc} we describe how the persistence of multifractal fluctuations can be monitored using the PDF of wavefunction intensities, and we proceed to revisit the generalized multifractal formalism. We present results from standard and multifractal finite-size scaling in Sections \ref{sec:sps} and \ref{sec:mfss}, including estimates for the critical parameters and multifractal exponents.
%
\section{The Anderson model in the unitary symmetry class}
\label{sec:model}
We consider the three-dimensional (3D) Anderson Hamiltonian in site basis, where time-reversal symmetry is explicitly broken by including random phases in the hopping terms,
\begin{equation}
\mathcal{H}=\sum_{k} \varepsilon_k \ket{k}\bra{k} - \sum_{\langle k,l\rangle} e^{i\phi_{kl}}\ket{k}\bra{l},
\label{eq:AH}
\end{equation}
where site $k=(x,y,z)$ is the position of an electron in a simple cubic lattice of linear size $L$ (measured in terms of the lattice constant), and $\langle k,l\rangle$ denote nearest neighbors. The random on-site energies $\varepsilon_k$ are uniformly distributed in the interval $[-W/2,W/2]$, and the random phases $\phi_{kl}$ are uniformly distributed in the range $[0,2\pi]$. In order for $\mathcal{H}$ to be Hermitian, we require that $\phi_{lk}=-\phi_{kl}$. The energy scale is set by the magnitude of the hopping elements, which is taken to be unity. Hamiltonian \eqref{eq:AH} may be viewed as that of a system in the presence of a random magnetic flux, yielding random Peierls phases for the hopping terms between neighboring lattice sites.\cite{Luttinger1951,Bohm2003}

Assuming periodic boundary conditions, the $L^3\times L^3$ Hamiltonian is diagonalized in the vicinity of $E=0$ (the center of the spectrum) for different linear sizes $L$ and degrees of disorder $W$, close to the critical value $W_c\simeq 18.8$ where the localization-delocalization transition occurs.\cite{Slevin2016}
For $W < W_c$ the system is in the delocalized (or metallic) phase while it is in the localized (or insulating) phase for $W > W_c$.
As the transition is approached, the localization (correlation) length $\xi$ of the eigenstates in the insulating (metallic) phase exhibits a power-law divergence, 
\begin{equation}
 \xi\propto|W-W_c|^{-\nu},
 \label{eq:xinu}
\end{equation}
where $\nu$ is the critical exponent determined by the universality class of $\mathcal{H}$.

Numerically, the eigenstates $\Psi=\sum_{j} \psi_j \ket{j}$ are obtained using the {\sc Jadamilu} library.\cite{Schenk2006a,Bollhofer2007} 
We consider only a single eigenstate per sample (disorder realization), namely, the eigenstate with energy closest to $E=0$. This is of primary importance in order to avoid the strong correlations that exist between eigenstates of the same sample.\cite{Rodriguez2011} 
Linear system sizes range from $20$ to $150$, and disorder values are in the interval $17.9\leqslant W \leqslant 19.7$. For each combination of size and disorder the number of disorder realizations varies from $2\times 10^4$ for the smallest $L$ to $10^4$ for the largest, adding up to a total of $\sim 4\,025\,000$ wavefunctions. The average number of states considered for each $L$-$W$ pair is indicated in Table \ref{tab:states}.

\begin{table}
\caption{Average number $\langle \mathcal{N}\rangle$ of uncorrelated wavefunctions generated for each choice of disorder $W$ and $L$. The maximum and minimum numbers of states for a given $W$  for each $L$ are shown in brackets. A total of 19 disorder values within the interval $[17.9,19.7]$ were considered.}
\begin{tabular}{cc}
\hline\hline
       $L$ & $\langle \mathcal{N} \rangle$ $(\mathcal{N}_{\rm max}, \mathcal{N}_{\rm min})$ \\
\hline
 20 & 20070 (20072, 20069) \\
 30 & 20393 (21127, 20081) \\
 40 & 20070 (20078, 20054) \\
 50 & 20049 (20050, 20048) \\
 60 & 20062 (20064, 20059) \\
 70 & 15149 (15225, 15104) \\
 80 & 15094 (15095, 15093) \\
 90 & 15089 (15090, 15088) \\
 100 & 15080 (15084, 15072) \\
 110 & 10047 (10048, 10047) \\
 120 & 10050 (10108, 10045) \\
 130 & 10043 (10044, 10039) \\
 140 & 10259 (11014, 10002) \\
 150 & 10349 (10864, 10011) \\
\hline\hline
\end{tabular}
\label{tab:states}
\end{table}
\section{Multifractality at the Anderson transition} 
\label{sec:mf}
\begin{figure}
  \includegraphics[width=\columnwidth]{\figdir/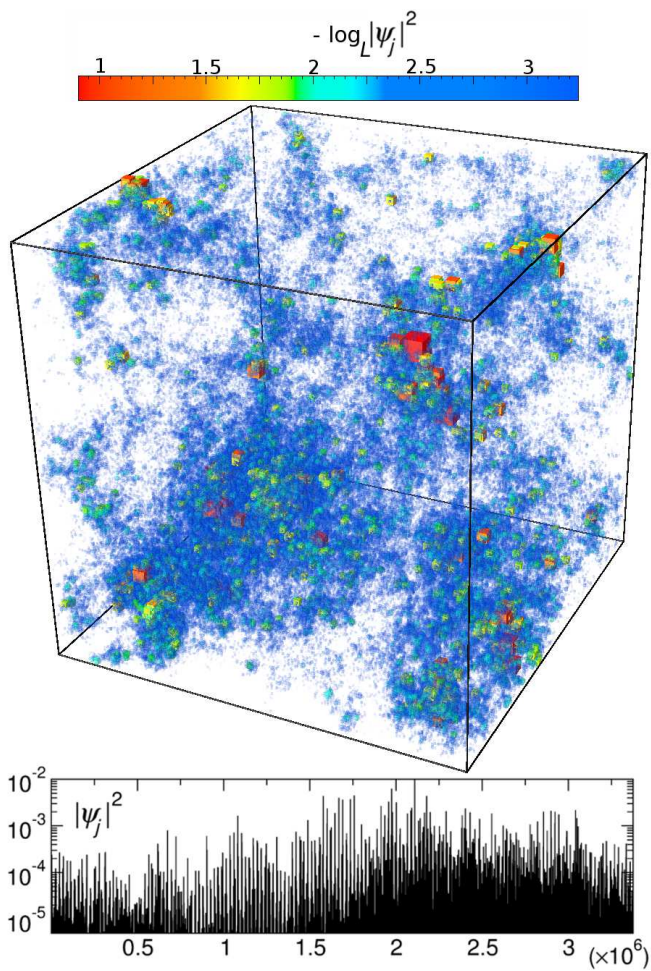}
  \caption{Critical eigenstate of Hamiltonian \eqref{eq:AH} near the band center ($E=0$) for $W=18.8$ and system size $L^3=150^3$. In the main plot, sites contributing to $98$\% of the norm of the wavefunction (omitting the sites with the lowest intensities) are shown as cubes whose volume is proportional to $|\psi_j|^2$.
The color and opacity of the cubes is chosen according to the value of $-\log_L|\psi_j|^2$, which ranges from $0.921$ to $3.231$. The bottom panel shows the wavefunction intensities versus site index $j\in[1,L^3]$.}
  \label{fig:MFstate}
\end{figure}

As first suggested by Aoki,\cite{Aoki1983,Aoki1986} the merging of the extended and localized characters of the wavefunction at the critical point of the Anderson transition would require the state to occupy an infinite volume as $L\rightarrow\infty$ ---like extended eigenstates do---, but at the same time a vanishing fraction of the whole system ---inheriting the non-ergodicity of localized states. This behavior is provided by a multifractal distribution.\cite{Mandelbrot2003,Paladin1987}
From an observational point of view, a multifractal wavefunction shows a pattern of large and intricate fluctuations of its intensity. 
A critical eigenstate of Hamiltonian \eqref{eq:AH} is shown in Fig.~\ref{fig:MFstate}. 

Mathematically, a multifractal eigenstate $\psi(\vb{r})$ is characterized by a power-law scaling of its moments,
\begin{equation}
 L^{d}\langle \overline{|\psi(\vb{r})|^{2q}}\rangle \sim L^{-\tau_q},
 \label{eq:MFmoments}
\end{equation}
where $d$ is the euclidean dimension of the system, the overline denotes a spatial average and the angular brackets mean a disorder average. The so-called mass exponents $\tau_q$ depend non-linearly on $q\in\mathbb{R}$.\cite{Janssen1994,Evers2008}  
For general $q$, we call the moments 
\begin{equation}
 R_q\equiv L^{d} \overline{|\psi(\vb{r})|^{2q}}=\int |\psi(\vb{r})|^{2q} \,\textrm{d}^d\vb{r}
\end{equation} 
{\em generalized inverse participation ratios} (GIPR).

Among the GIPR, the case $q=2$ corresponds to the standard {\em inverse participation ratio} (IPR), which measures the inverse of the subvolume of the system where the wavefunction has a noticeable amplitude, i.e.~it quantifies the spatial extension of the state.
For extended states it scales as $\text{IPR}_\text{metal}\sim L^{-d}$, while for localized states it will saturate as $L$ grows, $\text{IPR}_\text{insulator}\sim 1$.
The spatial extension of a multifractal state, however, scales as $\text{IPR}^{-1}\sim L^{\tau_2}$, where $\tau_2<d$, i.e.~its volume is unbounded but it occupies only a vanishing fraction of the whole system as $L\rightarrow\infty$.

In a multifractal wavefunction a set of points with the same $|\psi(\vb{r})|^2$ value, characterized by the variable 
\begin{equation}
\alpha(\vb{r})\equiv -\log_L |\psi(\vb{r})|^2,
\end{equation}
form a fractal with a certain fractal dimension $f(\alpha) \leqslant d$: The volume $V_\alpha$ of such a set, which corresponds to the number of points in a discrete lattice, scales as $V_\alpha\sim L^{f(\alpha)}$ as $L\rightarrow\infty$. The fractal dimension of the set depends on $\alpha$, i.e.~on the value of the wavefunction intensity. The whole collection of fractal dimensions occurring in the wavefunction is called the multifractal spectrum, $f(\alpha)$. 

It then ensues that $f(\alpha)$ is closely related to the PDF of the variable $\alpha$, i.e.~essentially to the PDF of the wavefunction intensities,\cite{Rodriguez2009}
\begin{equation}
 P_L(\alpha)\sim L^{f(\alpha)-d}.
 \label{eq:MFpdf}
\end{equation}

As for single fractals, multifractality persists at different length scales or under certain scale transformations, e.g.~after coarse-graining the distribution:
If we regularly partition the system into $(L/\ell)^d$ boxes of linear size $\ell$, and we integrate inside each box, 
\begin{equation}
\mu_k=\int_\text{box $k$}|\psi(\vb{r})|^2\,\textrm{d}^d\vb{r}
\end{equation}
or $\mu_k \equiv\sum_{j \in \text{box }k} |\psi_j|^2$ in a discrete lattice, the resulting distribution of 
integrated intensities defined on a system of linear size $L/\ell$ will retain the multifractal properties of 
the original wavefunction.
Thus, Eqs.~\eqref{eq:MFmoments}-\eqref{eq:MFpdf} hold true for the intensities $\mu_k$ upon the substitution $L\rightarrow L/\ell$, [e.g.~$\alpha_k\equiv-\ln \mu_k/\ln(L/\ell)$]. 
In this case the GIPR correspond to 
\begin{equation}
R_q=\sum_k \mu_k^q,
\end{equation}
which obey 
\begin{equation}
 \langle R_q \rangle \sim \lambda^{\tau_q}
 \label{eq:iprscal}
\end{equation}
in the limit that $\lambda\equiv\ell/L \rightarrow 0$.
Here, the brackets denote an \emph{ensemble} average over disorder.\cite{TypicalAv}

The general mathematical properties of a multifractal spectrum are well understood. \cite{Janssen1994,Evers2008} In particular, from Eqs.~\eqref{eq:MFmoments}--\eqref{eq:MFpdf} it follows that the function $f(\alpha)$ and the exponents $\tau_q$ are related by a Legendre transformation, 
\begin{equation}
    \alpha_q=d\tau_q/dq, \qquad f_q=q\alpha_q-\tau_q,
    \label{Legendre}
\end{equation}
which defines singularity strengths $\alpha_q$ and a singularity spectrum $f_q$.
The exponents $\tau_q$ are conveniently expressed in terms of anomalous scaling exponents $\Delta_q$,
\begin{equation}
\tau_q=d(q-1) +\Delta_q,
\end{equation}
which measure the deviation of the scaling of the GIPR from the metallic behavior, and determine the power-law nature of the spatial correlations of the multifractal wavefunctions. \cite{Evers2008,Delta2}
The anomalous scaling exponents are expected to obey a symmetry relation at the critical point,\cite{Mirlin2006,Symmetry}
\begin{equation}
  \Delta_q=\Delta_{1-q},
  \label{eq:Deltasym}
\end{equation}
which seems to hold for Anderson transitions in different systems and dimensionality,
\cite{Mildenberger2007,Mildenberger2007a,Evers2008a,Obuse2008,Obuse2007,Fyodorov2009,Rodriguez2008,Rodriguez2009,Monthus2009c} 
and has also been experimentally observed.\cite{Faez2009}

The study of the scaling of the GIPR with the length scales $L$ or $\ell$ constitutes the standard method to obtain numerically the multifractal spectrum.
\cite{Schreiber1991,Grussbach1995,Milde1997a,Vasquez2008,Rodriguez2008,Rodriguez2009a,Thiem2013a,Pinski2012}
\section{Multifractal fluctuations around the transition}
\label{sec:mffluc}
The analysis of the fundamental multifractal properties of the critical point requires knowledge of the position of the transition in the first place. Standard multifractal analysis  relies on this, and it is not useful in order to discern the existence (or absence) of the transition. The estimation of the position of the mobility edge and the critical exponent $\nu$, on the one hand, and the multifractal analysis, on the other, are so far completely decoupled. The persistence of multifractal fluctuations in the wavefunctions around the critical point\cite{Cuevas2007a} can however be used to bridge this gap and perform a full characterization of the transition.

\begin{figure}
  \includegraphics[width=.95\columnwidth]{\figdir/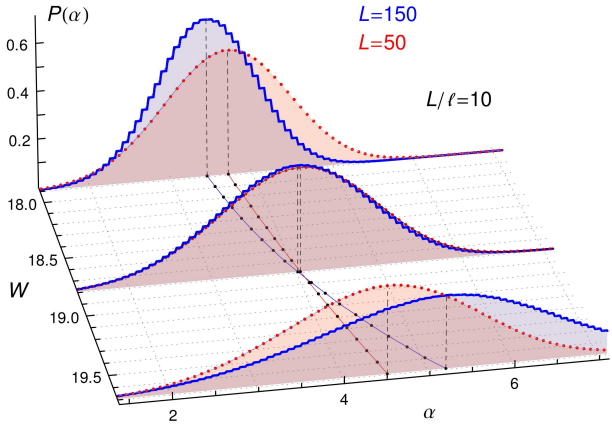}
  \caption{Scaling of the PDF of $\alpha$ across the disorder-induced metal-insulator transition, for different system sizes at fixed ratio $\lambda\equiv\ell/L=0.1$. The solid lines on the floor panel indicate the trajectories of the position of the PDF maximum versus $W$. The PDFs are obtained numerically, averaging over the total number of available wavefunctions (cp.~Table \ref{tab:states}).}
  \label{fig:pdf3D}
\end{figure}

\begin{figure*}[htb]
  \includegraphics[width=.66\textwidth]{\figdir/PDFflow.eps}\hfill
  \includegraphics[width=.31\textwidth]{\figdir/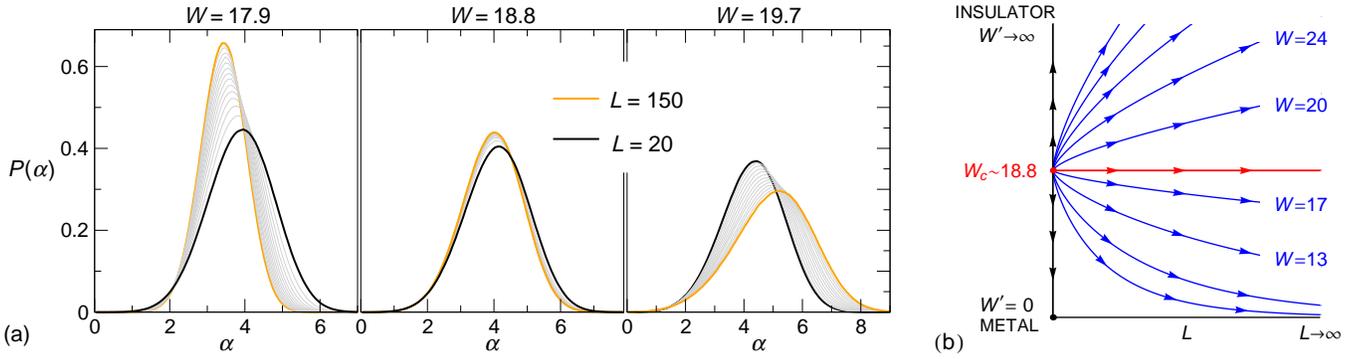}
  \caption{(a) Scaling of the PDF of $\alpha$ for $W=17.9$ (left, metallic phase), $W=18.8$ (middle, close to criticality), and $W=19.7$ (right, insulating phase), for system sizes $L\in[20,150]$ at fixed ratio $\lambda\equiv\ell/L=0.1$. (b) Qualitative schematic flow of the coarse-graining transformation of the wavefunction with box size $\ell=\lambda L$ for fixed $\lambda$, assuming $\nu>1$. The vertical axis corresponds to the renormalized disorder $W'$ and the horizontal axis to the system size $L$. For a given disorder $W$ the flow follows the corresponding blue line as $L$ grows. Close to the fixed point $W_c$ the renormalized disorder obeys Eq.~\eqref{eq:RGdis}. }
  \label{fig:flowpdfs}
\end{figure*}

The starting point is the multifractal scaling of the PDF for the integrated distribution $\mu_k$, 
\begin{equation}
 P_{L/\ell}(\alpha)\sim \left(\frac{L}{\ell}\right)^{f(\alpha)-d}.
 \label{eq:PDF}
\end{equation}
This relation implies that at the critical point the only relevant length scale in the PDF is the ratio $\lambda=\ell/L$, thus different system sizes will exhibit the same distribution of intensities when the box size $\ell$ is appropriately chosen. Away from the critical point, however, this is not true, since a length scale independent $f(\alpha)$ does not exist (no strict multifractality), and the PDF must contain additional functional dependencies on $L$ and $\ell$.\cite{Rodriguez2010} 
This is clearly demonstrated in Figs.~\ref{fig:pdf3D} and \ref{fig:flowpdfs}(a), where the flow of the numerically obtained PDF for our system is shown as a function of disorder and system size.%
At the critical point the whole distribution is invariant upon changing the system size ---up to finite-size irrelevant scaling corrections. As one moves away from the transition the distribution becomes $L$-dependent again,
exhibiting a standard scaling behavior. 
To the best of our knowledge, such a study of the PDF in the absence of time-reversal symmetry has not been previously reported.

The coarse-graining of the wavefunction acts as a renormalization transformation of all length scales in the system by $\ell=\lambda L$, for constant $\lambda$.\cite{Cardy1996} 
In the insulating phase the localization length will then transform as 
\begin{equation}
 \xi'=\xi/\ell\sim \xi/L.
 \label{eq:RGxi}
\end{equation}
Therefore, upon coarse-graining a localized state becomes more localized as $L$ increases; consequently the frequency of low intensities grows, and $P(\alpha;W,L,\ell)$ shifts towards larger $\alpha$ values. Similarly, Eq.~\eqref{eq:RGxi} also applies to the correlation length in the metallic phase: Upon renormalization, an extended state becomes more and more homogeneous with increasing $L$, and the PDF moves towards the limiting form $P_\text{metal}(\alpha)\underset{L\rightarrow\infty}{=}\delta(\alpha-d)$.
From Eqs.~\eqref{eq:xinu} and \eqref{eq:RGxi}, it follows that the degree of disorder renormalizes as 
\begin{equation}
  |W'-W_c|\sim |W-W_c|L^{1/\nu}, 
  \label{eq:RGdis}
\end{equation}
where $W_c$ is the fixed point of the transformation. 
Thus, disorder renormalizes to larger (smaller) values in the insulating (metallic) phase as $L$ grows. 
A schematic plot $W'$ versus $L$ of the flow of the coarse-graining transformation is shown in Fig.~\ref{fig:flowpdfs}(b). 

The shown scaling of the PDF implies that just from histograms of intensities and observing how these behave under the described length scale transformation, it is possible to unambiguously identify a localization-delocalization transition.\cite{PDF} 
It also provides an alternative interpretation of the Anderson transition as the fixed point of the coarse-graining transformation of the wavefunctions. In particular, we believe that this approach would be valuable to analyze experimental data, such as local density of states measurements obtained by scanning tunnelling microscopy on solid state devices.
\cite{Morgenstern2002,Hashimoto2008,Richardella2010,Miller2010,Krachmalnicoff2010,Hashimoto2012,Bindel2017}
\subsection{Generalized multifractal scaling}
\label{sec:gmfs}
This qualitative scaling picture translates into a quantitative analysis upon formulating appropriate scaling laws for the relevant quantities around the critical point. 
In general, the PDF will depend on $W$, $L$, and $\ell$ away from the critical point: $P(\alpha;W, L, \ell)$. The behavior under the renormalization transformation at fixed $\lambda=\ell/L$ described above suggests that close to the critical point the function can be written as 
\begin{equation}
 \mathcal{P}(\alpha;|W-W_c| L^{1/\nu}, L/\ell)=\widehat{\mathcal{P}}(\alpha;L/\xi,L/\ell),
\end{equation}
or in the entirely equivalent form $\widetilde{\mathcal{P}}(\alpha;L/\xi,\ell/\xi)$. We then assume that around the transition relevant quantities are determined by the ratios of the length scales $\ell$ and $L$ to the localization (correlation) length. This statement is in fact the underlying basis for the scaling theory of localization.
\cite{Abrahams1979,Lee1985a}

We proceed to define a {\em generalized multifractal analysis} valid close to the critical point. 
The scaling law for the GIPR reads \cite{Yakubo1998}
\begin{equation}
  \langle R_q\rangle (W, L, \ell) = \lambda^{\tau_q} \mathcal{R}_q(L/\xi,\ell/\xi),
  \label{eq:GIPRscaling}
\end{equation}
which can be rearranged introducing {\em generalized mass exponents} 
\begin{equation}
 \widetilde{\tau}_q(W,L,\ell) \equiv \frac{\ln \langle R_q\rangle (W, L, \ell)}{\ln\lambda}, 
\end{equation}
obeying 
\begin{equation}
 \widetilde{\tau}_q(W,L,\ell) = \tau_q + \frac{q(q-1)}{\ln\lambda}\mathcal{T}_q (L/\xi,\ell/\xi).
 \label{eq:GMFESscaling}
\end{equation}
Here, the tilde is used to emphasize that this equation applies throughout the critical region and not just at the critical point.
Similarly one can define {\em generalized anomalous scaling exponents} and {\em generalized singularity strengths},  
\begin{align}
 \widetilde{\Delta}_q &\equiv \widetilde{\tau}_q-d(q-1), \\
 \widetilde{\alpha}_q &\equiv d \widetilde{\tau}_q/dq = \big\langle \sum_k \mu_k^q \ln\mu_k \big\rangle/ \left(\langle R_q \rangle \ln \lambda\right),
 \label{eq:def-alphaq}
\end{align}
which obey scaling laws similar to Eq.~\eqref{eq:GMFESscaling}.

The {\em generalized multifractal exponents} (GMFE) become the usual scale invariant multifractal exponents at the critical point $W_c$ in the limit $L/\ell\rightarrow\infty$ ($\lambda\rightarrow 0$). For more details about the GMFE we refer the reader to Ref.~\onlinecite{Rodriguez2011}.

Fitting the variation of the GMFE with disorder, system size and box size using scaling laws of the form \eqref{eq:GMFESscaling} allows for
the estimation of the $q$-independent critical parameters $W_c$ and $\nu$, and the simultaneous determination of a multifractal exponent for a particular $q$.
We call this approach {\em multifractal finite-size scaling} (MFSS).
\section{Single parameter scaling at fixed $\boldsymbol{\lambda}$}
\label{sec:sps}
The scaling function in Eq.~\eqref{eq:GIPRscaling} can be written equivalently as $\mathcal{R}_q(L/\xi,\lambda)$. This form suggests that a standard single parameter finite-size scaling (FSS) procedure is applicable by considering data at a fixed value of $\lambda$, which does not, however, permit the estimation of the scale invariant multifractal exponents.
In this case the scaling laws for the GMFE become one parameter functions, 
\begin{equation}
 \Gamma_q(W,L)= \mathcal{G}_q(L/\xi),
 \label{eq:gen}
\end{equation}
where $\Gamma_q$ denotes any of the above mentioned exponents.

In order to fit data for the GMFE, we follow a standard procedure and include two
kinds of corrections to scaling:\cite{Slevin1999,Slevin2014}
(i) nonlinearities of the $W$ dependence of the scaling variables,
and (ii) an irrelevant scaling correction that accounts for a shift with $L$ of the apparent critical disorder at which the $\Gamma_q(W,L)$ curves cross.
After expanding to first order in the irrelevant scaling term,
the scaling functions take the form
\begin{equation}
  \mathcal{G}_q(\varrho L^{1/\nu}, \eta L^{y})= {\mathcal G}_q^0(\varrho L^{1/\nu}) +\eta L^{y} {\mathcal G}_q^1(\varrho L^{1/\nu}).
  \label{eq:Gexpansion}
\end{equation}
Here $\varrho$ and $\eta$ are the relevant and irrelevant scaling variables, respectively.
The irrelevant component is expected to vanish for large $L$, so $y<0$.
Both scaling functions are Taylor-expanded
\begin{equation}
 \mathcal{G}_q^k(\varrho L^{1/\nu})= \sum_{j=0}^{n_k} a_{kj}\varrho^j L^{j/\nu}, \quad \text{for } k=0, 1.
\end{equation}
The scaling variables are expanded in terms of $w\equiv(W-W_c)$
to order $m_{\varrho}$ and $m_{\eta}$, respectively,
\begin{equation}
  \varrho(w)=w+\sum_{m=2}^{m_{\varrho}} b_m w^m,\quad
   \eta(w)=1+\sum_{m=1}^{m_{\eta}} c_m w^m.
   \label{eq:fieldex}
\end{equation}
The fitting function is characterized by the expansion orders $n_0, n_1, m_{\varrho}, m_{\eta}$.
The total number of free parameters to be determined in the fit is
$N_P=n_0+ n_1+m_{\varrho}+ m_{\eta}+4$ (including $\nu$, $y$ and $W_c$).

The localization (correlation) length, up to a constant of proportionality, is $\xi=|\varrho(w)|^{-\nu}$.
After subtraction of corrections to scaling,
\begin{equation}
\Gamma_q^\text{corr}\equiv\Gamma_q(W,L)- \eta L^{y} {\mathcal G}_q^1(\varrho L^{1/\nu}),
\end{equation}
the data for the GMFE should fall on the single-parameter curves
\begin{equation}
\Gamma_q^\text{corr} = \mathcal{G}_q^0(\pm (L/\xi)^{1/\nu}).
\end{equation}
\subsection{Numerical procedure and results}
\label{sec:spsresults}
When performing FSS, the aim is to identify a \emph{stable} expansion of the scaling function that fits the numerical data.
The best fit is found by minimizing the $\chi^2$ statistic over the parameter space.
The validity of the fit is decided by the $p$-value or goodness-of-fit.
We take $p\geqslant 0.1$ as the threshold for an acceptable fit.
As a rule of thumb the expansion orders $n_0, n_1, m_{\varrho}, m_{\eta}$ are kept as low as possible while giving acceptable and stable fits.
Once a stable fit has been found, the precision of the estimates of the critical parameters is estimated by a Monte Carlo simulation, i.e.\ by fitting a large set of synthetic data sets generated by adding appropriately scaled random normal errors to an ideal data set generated from the best-fit model.
For a detailed description of the FSS procedure we refer the reader to the Appendices provided in Ref.~\onlinecite{Rodriguez2011}.

\begin{figure}
 \centering
 \includegraphics[width=.9\columnwidth]{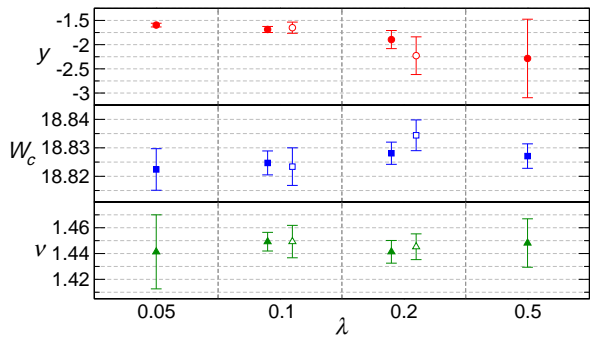}
 \caption{The estimates of the critical disorder $W_c$, critical exponent $\nu$, and irrelevant exponent $y$, obtained from single parameter FSS at fixed $\lambda$ (see Table \ref{tab:lamfits}). Error bars are 95\% confidence intervals. Full (empty) symbols show results from the scaling of $\widetilde{\alpha}_0$ ($\widetilde{\alpha}_1$).}
 \label{fig:CRITlam}
\end{figure}

We performed a FSS analysis for $\widetilde{\alpha}_0$ and $\widetilde{\alpha}_1$ for different values of $\lambda$ ranging from 0.05 to 0.5. 
All relevant details of the fits together with the estimated critical parameters are given in Table \ref{tab:lamfits}. The obtained values for $y$, $W_c$ and $\nu$ are shown as functions of $\lambda$ in Fig.~\ref{fig:CRITlam}. 
\begin{table*} 
\caption{The estimates of the critical parameters together with 95\% confidence intervals,
from single parameter finite-size scaling at fixed $\lambda$ of $\widetilde{\alpha}_0$ and $\widetilde{\alpha}_1$, under ensemble average.
The number of data points used is $N_D$ (average percentage precision in parentheses),
the number of free parameters in the fit is $N_P$, $\chi^2$ is the value of the chi-squared statistic for the best fit, and $p$ is the goodness-of-fit probability. The last column specifies the orders of the expansion. The system sizes considered are $L\in[20, 150]$ and the range of disorder is  $W\in\left[17.9,19.7\right]$.}
\label{tab:lamfits}
\begin{tabular}{cccccccccc}
\hline\hline
   GMFE & $\lambda$  & $\nu$  & $W_c$ & $-y$ & $N_D$(prec.\%) & $N_P$ & $\chi ^2$ & $p$ & $n_0$ $n_1$ $m_\rho$ $m_\eta$ \\\hline
 $\widetilde{\alpha}_0$ & 0.05 & 1.441(1.417,1.470) & 18.822(18.814,18.830) & 1.60(1.56,1.63) & 133(0.03) & 12 & 139 & 0.13 & 4\,2\,1\,1 \\
 $\widetilde{\alpha}_0$ & 0.1  & 1.449(1.442,1.456) & 18.825(18.820,18.829) & 1.69(1.62,1.75) & 266(0.04) & 11 & 257 & 0.45 & 4\,2\,1\,0 \\
 $\widetilde{\alpha}_0$ & 0.2  & 1.441(1.433,1.450) & 18.828(18.824,18.832) & 1.90(1.71,2.08) & 266(0.06) & 11 & 254 & 0.51 & 4\,2\,1\,0 \\
 $\widetilde{\alpha}_0$ & 0.5  & 1.448(1.430,1.467) & 18.827(18.821,18.831) & 2.29(1.53,3.10) & 266(0.10) & 12 & 225 & 0.91 & 4\,2\,1\,1 \\
 $\widetilde{\alpha}_1$ & 0.1  & 1.449(1.437,1.462) & 18.823(18.817,18.830) & 1.65(1.54,1.77) & 266(0.10) & 11 & 280 & 0.13 & 4\,2\,1\,0 \\
 $\widetilde{\alpha}_1$ & 0.2  & 1.445(1.436,1.455) & 18.834(18.828,18.840) & 2.23(1.87,2.62) & 266(0.11) & 11 & 273 & 0.21 & 4\,1\,2\,0 \\\hline\hline
\end{tabular}
\end{table*}
\begin{figure*}
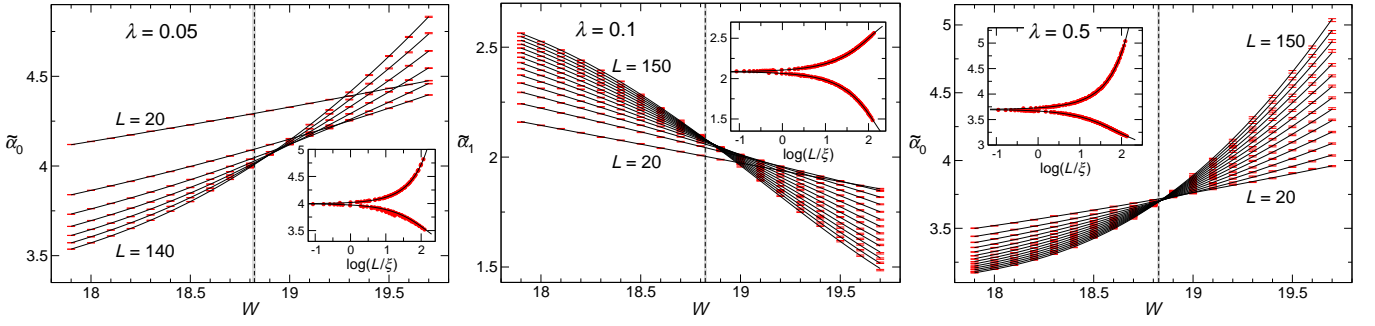

 \includegraphics[width=.33\textwidth]{\figdir/FITalpha0l005.eps}\hfill
 \includegraphics[width=.33\textwidth]{\figdir/FITalpha1l01.eps}\hfill
 \includegraphics[width=.33\textwidth]{\figdir/FITalpha0l05.eps}
 \caption{Plots of the GMFE $\widetilde{\alpha}_0$ and $\widetilde{\alpha}_1$ for $\lambda=0.05$ (left), $\lambda=0.1$ (middle), and $\lambda=0.5$ (right) as functions of disorder at various system sizes $L\in[20,150]$. The error bars are standard deviations. The lines are the best fits listed in Table \ref{tab:lamfits}.
 The estimated $W_c$ are shown by vertical dashed lines and 95\% confidence intervals by the shaded regions.
 The insets show the data plotted versus $L/\xi$ with the irrelevant contribution subtracted and the scaling function (solid line).}
 \label{fig:FSSlam}
\end{figure*}

The values of the critical disorder and critical exponent resulting from FSS are mutually consistent for different $\lambda$ values and both $q=0$ and $q=1$. The average precision of the numerical data for the GMFE degrades slowly when $\lambda$ grows (see Table \ref{tab:lamfits}), since larger $\lambda$ means smaller renormalized system sizes, and in general broader distributions of the numerical GMFE. Interestingly, the amplitude of the irrelevant scaling contribution also depends strongly on $\lambda$, being largest for small $\lambda$ values (assuming a fixed set of available system sizes). This is clearly seen in Fig.~\ref{fig:FSSlam} which shows the FSS fits of the numerically calculated  $\widetilde{\alpha}_0$ and $\widetilde{\alpha}_1$. (This dependence of the irrelevant correction on $\lambda$ has not been expliclity observed in previous FSS studies in the unitary symmetry class,\cite{Ujfalusi2015a} and it is entirely consistent with our ansatz for the MFSS scaling function in Sec.~\ref{sec:mfss}.) This strong variation of the irrelevant amplitude induces some fluctuation in the estimated irrelevant exponent and consequently in the position of the critical point. Nevertheless, the estimates for $\nu$ are remarkably stable over the $\lambda$-range considered. Since we average over a large number of samples, even the data for $\widetilde{\alpha}_0$ for $\lambda=0.5$ is accurate enough to provide stable fits and reasonable confidence intervals for the estimated critical parameters. Note that at the latter $\lambda$ value the coarse-grained wavefunction lives in a renormalized system of size $L'=2$. 

The resulting estimates for $\nu$ and $W_c$ are compatible with previous transfer-matrix calculations for the Anderson transition in the presence of a random magnetic flux.\cite{Kawarabayashi1998a, Kawarabayashi1998, Slevin2016}
\section{Multifractal finite-size scaling}
\label{sec:mfss}
Now we exploit the full potential of scaling laws of the form \eqref{eq:GMFESscaling}, fitting the variation of the GMFE as function of disorder $W$, system size $L$ and box size $\ell$. This leads to a \emph{simultaneous} estimation of the multifractal exponents and the critical parameters, $\nu$, $W_c$, and $y$.

Besides the more involved form of the scaling functions, a proper MFSS needs to take data correlations into account: 
Different coarse-graining $\ell$ for the same disorder $W$ and system size $L$ use the same set of wavefunctions,
leading to correlated estimates of the GMFE for different $\ell$ and the same $W$ and $L$.
We generalize the definition of $\chi^2$ in the numerical minimization by including the full covariance matrix for the GMFE.
A detailed description of the calculation of the covariance matrix and the $\chi^2$-minimization procedure is given in Ref.~\onlinecite{Rodriguez2011}.

In MFSS, the scaling functions include two variables, $L/\xi$ and $\ell/\xi$, which can vary independently but renormalize in the same way.
As for FSS, we need to allow for a non-linear dependence in $W$ and for irrelevant scaling variables. 
In agreement with the behavior observed in the orthogonal symmetry class,\cite{Rodriguez2011} here we also find that the most important irrelevant contribution is due to the box size $\ell$. Therefore, we use the expansion
\begin{multline}
  \widetilde{\Delta}_q(\varrho L^{1/\nu}, \varrho \ell^{1/\nu}, \eta \ell^{y}) = \\ \Delta_q+\frac{1}{\ln (\ell/L)} \sum_{k=0}^2 \left(\eta \ell^{y}\right)^k \mathcal{T}^k_q(\varrho L^{1/\nu}, \varrho \ell^{1/\nu}),
  \label{eq:expand3D}
\end{multline}
for the generalized anomalous scaling exponent, and similarly for $\widetilde{\alpha}_q$.
Here, $\varrho$ and $\eta$ are the relevant and irrelevant scaling variables, with $1/\nu$ and $y<0$ the corresponding exponents.
In order to maximize the amount of data and the range of box sizes which we can fit reliably, we expand to second order in the irrelevant variable.
This is in contrast to the study of Ref.~\onlinecite{Ujfalusi2015a}, where the expansion in the irrelavant variable was restricted to first order and hence only considerably smaller data sets could be fitted.

The functions $\mathcal{T}_q^k$ are expanded,
\begin{equation}
 \mathcal{T}^k_q(\varrho L^{1/\nu}, \varrho \ell^{1/\nu})=\sum_{i=0}^{n_L^k} \sum_{j=0}^{n_\ell^k} a_{kij} \varrho^{i+j} L^{i/\nu} \ell^{j/\nu},
\end{equation}
for $k=0,1,2$, as are the scaling variables [see Eq.~\eqref{eq:fieldex}].
The expansion of the scaling function is then characterized by the indices $n^0_L,n^0_\ell,n^1_L,n^1_\ell,n^2_L,n^2_\ell,m_\varrho,m_\eta$.
[In order to consider a most general fit, we change independently the expansion orders of the two relevant variables in $\mathcal{T}_q^k$ (cf.~Ref.~\onlinecite{Ujfalusi2015a}).]
The number of free parameters is given by
\begin{equation}
N_P=\sum_{k=0}^2 (n_L^k+1)(n_\ell^k+1) +m_\varrho +m_\eta +3.
\end{equation}
After subtraction of irrelevant corrections we have
\begin{equation}
\widetilde{\Delta}_q^\text{corr}=\Delta_q+ \mathcal{T}^0_q(\pm (L/\xi)^{1/\nu}, \pm (\ell/\xi)^{1/\nu})/\ln(\ell/L)
\end{equation}
and the numerical data should fall on a common scaling surface.

Note that, when evaluated at fixed $\lambda$, Eq.~\eqref{eq:expand3D} leads to the FSS expansion considered in 
Sec.~\ref{sec:sps}, where the amplitude of the irrelevant terms is proportional to $\lambda^{-|y|}$, and thus grows when $\lambda$ decreases, as observed in Fig.~\ref{fig:FSSlam}.
%
\subsection{Results}
\label{sec:mfssresults}
For the MFSS analysis we considered the ensemble averaged GMFE $\widetilde{\Delta}_q$ for different $q\in[-1,2]$, and $\widetilde{\alpha}_q$ for $q=0,1$.
The estimates of the critical parameters and the multifractal exponents, together with full details of the fits, are included in Table \ref{tab:results3D}.
\begin{table*} 
\caption{The estimates of the critical parameters and multifractal exponents together with 95\% confidence intervals,
from MFSS of $\widetilde{\Delta}_q$ for $q\in[-1,2]$ and $\widetilde{\alpha}_q$ for $q=0,1$, under ensemble average.
The number of data points used is $N_D$ (average percentage precision in parentheses),
the number of free parameters in the fit is $N_P$, $\chi^2$ is the value of the chi-squared statistic for the best fit, and $p$ is the goodness-of-fit probability. The last column specifies the orders of the expansion: $n^0_L,n^0_\ell,n^1_L,n^1_\ell,n^2_L,n^2_\ell,m_\varrho,m_\eta$. The system sizes considered are $L\in[20, 150]$, the range of disorder is  $W\in\left[18.1,19.5\right]$, minimum box size $\ell_{\rm min}=2$ $(\lambda_{\rm min}=0.013)$ for $q\leqslant0.75$ and $\ell_{\rm min}=1$ $(\lambda_{\rm min}=0.0067)$ for $q\geqslant 1$. The maximum values considered for $\lambda$ change from $\lambda_{\rm max}=0.045$ to $\lambda_{\rm max}=0.1$ for different $q$. For $q\geqslant 1$ and $L\geqslant 130$ the disorder range is reduced to $W\in\left[18.5,19.1\right]$.%
}
\label{tab:results3D}
\begin{tabular}{cccccccccc}
\hline\hline
$q$ & $\Delta_q$ ($\alpha_q$ for $q=0,1$) & $\nu$  & $W_c$ & $-y$ & $N_D$(prec.) & $N_P$ & $\chi^2$ & $p$ & Expansion \\\hline 
 $-$1    & $-$1.9512($-$1.9545,$-$1.9480) & 1.442(1.433,1.451) & 18.824(18.821,18.827) & 1.848(1.826,1.871) & 840(0.17) & 19 & 842 & 0.30 & 3\,1\,1\,1\,0\,1\,1\,1 \\
 $-$0.75 & $-$1.3239($-$1.3254,$-$1.3224) & 1.444(1.436,1.452) & 18.825(18.822,18.827) & 1.841(1.827,1.855) & 840(0.14) & 22 & 813 & 0.54 & 3\,2\,0\,1\,0\,1\,2\,1 \\
 $-$0.5  & $-$0.7812($-$0.7818,$-$0.7806) & 1.446(1.440,1.451) & 18.823(18.821,18.825) & 1.835(1.825,1.844) & 840(0.13) & 21 & 812 & 0.57 & 2\,2\,0\,2\,0\,1\,2\,2 \\
 $-$0.25 & $-$0.3356($-$0.3358,$-$0.3353) & 1.445(1.439,1.450) & 18.824(18.823,18.826) & 1.827(1.819,1.835) & 840(0.13) & 22 & 825 & 0.43 & 3\,2\,0\,1\,0\,1\,2\,1 \\
  0    &  4.1004(4.0994,4.1013)   & 1.446(1.440,1.452) & 18.824(18.822,18.826) & 1.808(1.800,1.817) & 840(0.04) & 22 & 840 & 0.29 & 3\,2\,0\,1\,0\,1\,2\,1 \\
 0.25  &  0.2096(0.2094,0.2099)   & 1.446(1.436,1.457) & 18.823(18.821,18.825) & 1.783(1.770,1.797) & 630(0.13) & 23 & 646 & 0.13 & 3\,2\,0\,2\,0\,2\,1\,1 \\
 0.5   &  0.2809(0.2805,0.2813)   & 1.450(1.441,1.458) & 18.821(18.819,18.824) & 1.765(1.741,1.788) & 540(0.13) & 23 & 553 & 0.13 & 3\,2\,0\,2\,0\,2\,2\,0 \\
 0.75  &  0.2092(0.2087,0.2097)   & 1.448(1.436,1.461) & 18.823(18.820,18.827) & 1.758(1.709,1.806) & 540(0.15) & 27 & 552 & 0.12 & 4\,2\,0\,2\,0\,2\,3\,0 \\
  1    &  1.9122(1.9056,1.9193)   & 1.454(1.440,1.468) & 18.814(18.806,18.823) & 1.590(1.526,1.654) & 540(0.07) & 30 & 559 & 0.07 & 4\,3\,0\,2\,0\,2\,1\,0 \\
 1.25  & $-$0.3317($-$0.3343,$-$0.3288) & 1.457(1.439,1.476) & 18.822(18.809,18.834) & 1.625(1.523,1.728) & 540(0.15) & 30 & 551 & 0.10 & 6\,2\,0\,1\,0\,1\,2\,0 \\
 1.5   & $-$0.7618($-$0.7663,$-$0.7575) & 1.441(1.419,1.464) & 18.819(18.807,18.831) & 1.567(1.485,1.655) & 661(0.18) & 30 & 688 & 0.06 & 6\,2\,0\,1\,0\,1\,2\,0 \\
 1.75  & $-$1.2690($-$1.2776,$-$1.2603) & 1.437(1.420,1.453) & 18.814(18.798,18.830) & 1.513(1.393,1.638) & 661(0.21) & 29 & 683 & 0.08 & 4\,3\,1\,0\,1\,0\,2\,0 \\
  2    & $-$1.8347($-$1.8490,$-$1.8197) & 1.435(1.400,1.465) & 18.816(18.794,18.836) & 1.536(1.355,1.738) & 661(0.25) & 31 & 675 & 0.10 & 4\,2\,3\,1\,2\,0\,2\,0 \\\hline\hline
\end{tabular}
\end{table*}
\begin{figure*}
 \includegraphics[height=6.1cm]{\figdir/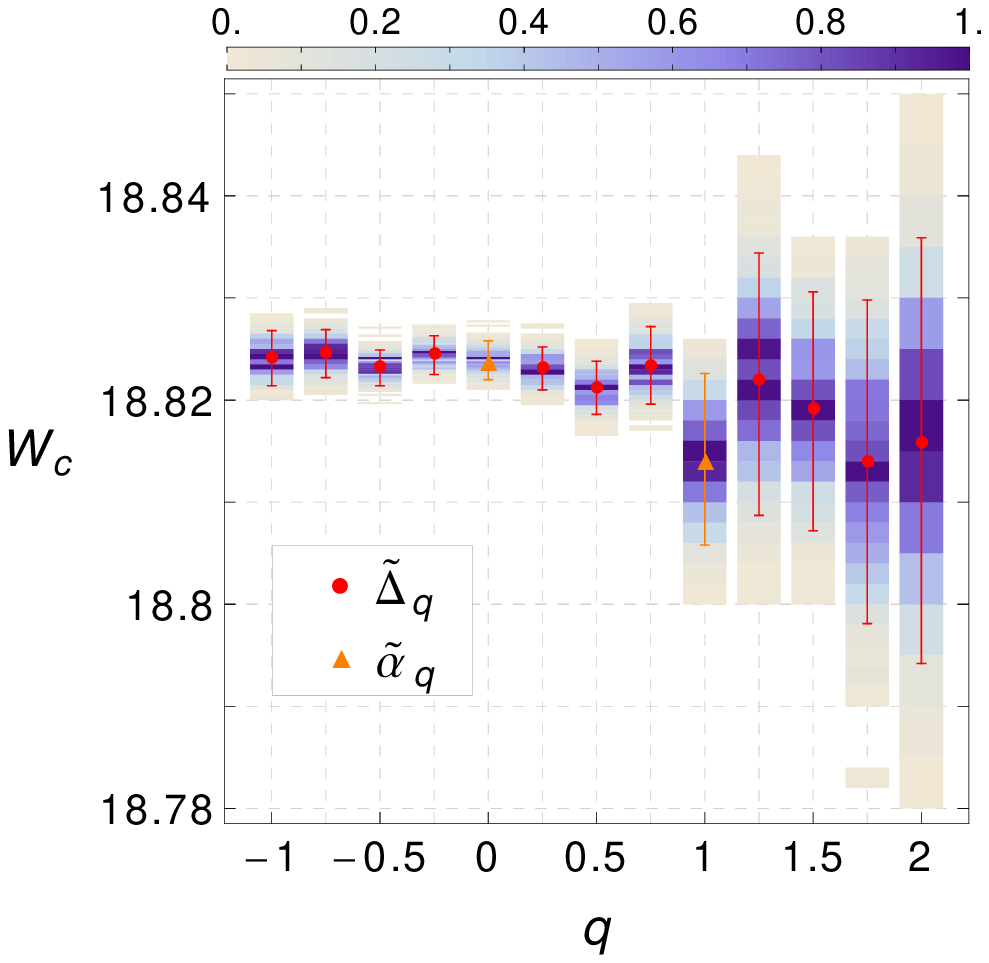}\hfill
 \includegraphics[height=6.1cm]{\figdir/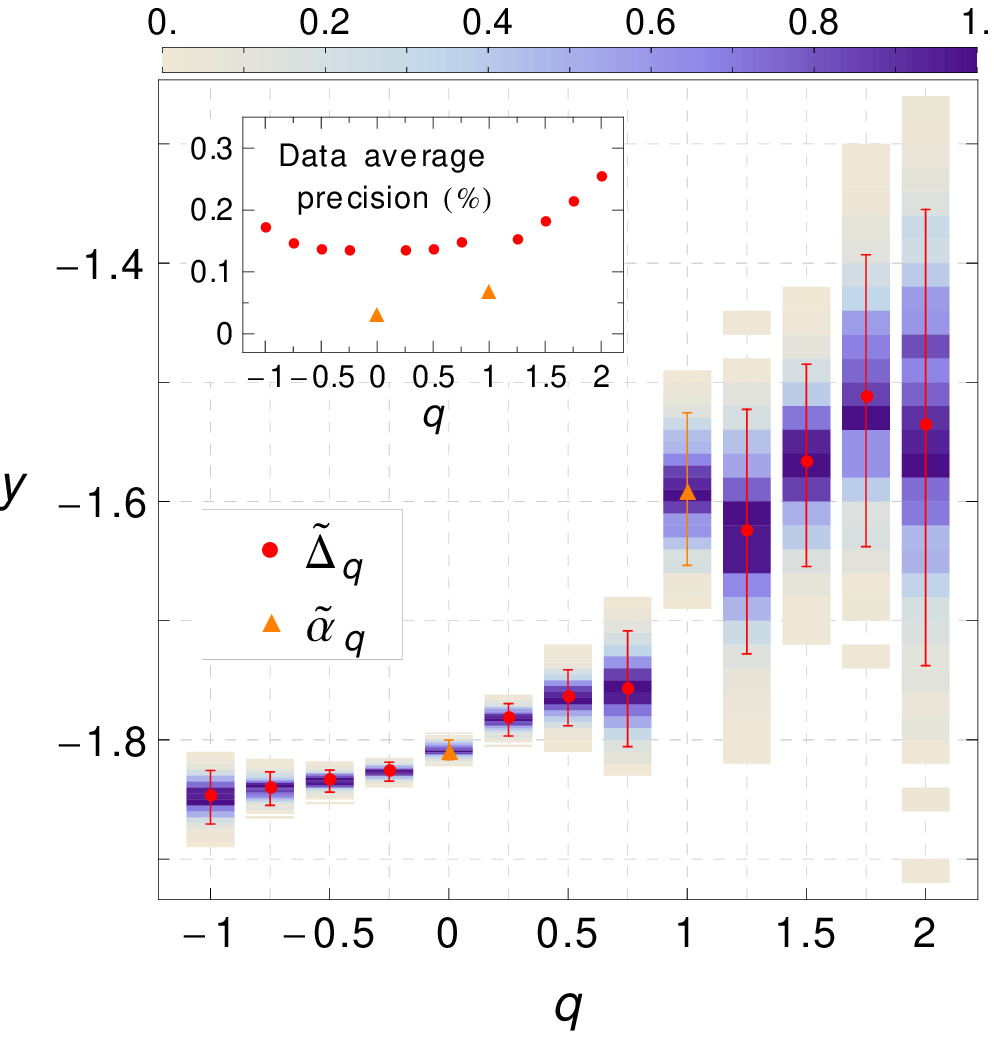}\hfill
 \includegraphics[height=6.1cm]{\figdir/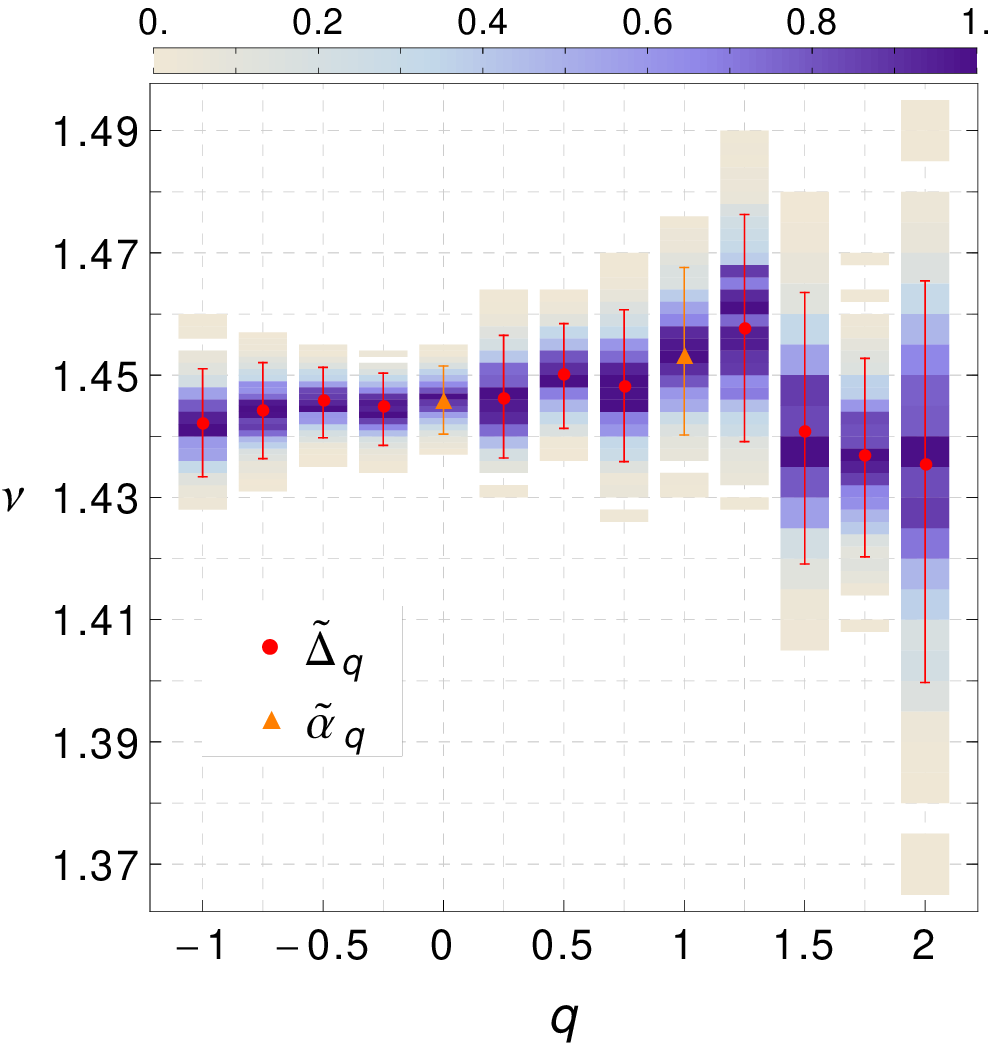}
 \caption{The estimates of the critical parameters $W_c$, $y$, and $\nu$, as functions of $q$, obtained from MFSS for $\widetilde{\Delta}_q$ and $\widetilde{\alpha}_q$ (only $q=0,1$). Error bars are 95\% confidence intervals. The corresponding values are listed in Table \ref{tab:results3D}. The inset in the middle plot shows the average data precision versus $q$ for the data set used. A density plot of the histograms obtained from the Monte Carlo simulations used to determine the uncertainty of the estimates is shown for each $q$. The color scale on top of each graph is for the density plot. The histograms are normalized so that their maximum value is unity.}
 \label{fig:CPvsq3D}
\end{figure*}

For each $q$ value we considered a different range of data, trying to maximize the number of points that we could fit reliably.
For negative and small $q$ values, we exclude the case $\ell=1$ in order to minimize the appearance of errors induced by inaccuracies in the small amplitudes of the calculated eigenstates. The minimum values of $\lambda$ included in the data sets are $\lambda_{\rm min}=0.013$ ($\ell_{\rm min}=2$) for $q\leqslant0.75$ and $\lambda_{\rm min}=0.0067$ ($\ell_{\rm min}=1$) for $q\geqslant 1$. 

The best-fit estimates for the critical parameters  as functions of $q$ are shown in Fig.~\ref{fig:CPvsq3D}.
We obtain a remarkable consistency of the estimates for $W_c$ and $\nu$, which must be $q$ independent. 
There is, however, an apparent fluctuation of the estimated value of the irrelevant exponent $y$, whose magnitude shifts to smaller values for $q\geqslant 1$. This is correlated to the fact that data for $\ell=1$ ---which exposes the largest irrelevant corrections--- is considered for these $q$ values. On the other hand, we emphasize that there is in principle no reason why the irrelevant terms and $y$ should be independent of $q$. 

For $q\leqslant0$ we succeed in fitting reliably remarkably large data sets using a reasonable number of parameters in the scaling function. The resulting estimates of the critical parameters from this $q$-range are very stable and exhibit low uncertainty. Data for $q\geqslant 0.75$, however, turned out to be more challenging: The data sets that can be reliably fit are smaller, and higher order expansions (and consequently more parameters) are required. In turn, this translates into larger uncertainties for $W_c$ and $\nu$. 
This behavior is to be expected: For $q\gtrsim 1$, the interval between the metallic and insulating limits for the values of the GMFE, $\widetilde{\Delta}_{q\geqslant1}\in[0,-d(q-1)]$, is considerably reduced when compared to the case $q\leqslant0$, where $\widetilde{\Delta}_{q\leqslant0}\in[0,-\infty]$ (cp.~Fig.~2 in Ref.~\onlinecite{Rodriguez2011}). This implies that within the same $W$-range, the curvature of the data for $q\gtrsim 1$ will be higher as $L$ increases, since the metallic and insulating bounds close in faster. 

Additionally, in order to maximize the number of fitted data for $q\geqslant1$ we relaxed the goodness-of-fit criterion and regarded any fit with $p\geqslant0.05$ as acceptable. We find that this helps to reduce the ambiguity of stable fits in this $q$-range. This choice for the $p$-threshold may indeed be well justified, since the uncertainty of the data in this range might be slightly underestimated. By increasing the number of disorder realizations the uncertainty of the numerically obtained GMFE could in principle be continuously reduced. The data, however, is also affected by the error of the numerically obtained wavefunction amplitudes, which has been so far ignored. This latter error establishes a lower bound for the uncertainty of the averaged GMFE: If the mean relative error in the wavefunction amplitudes is $\sigma_\psi$, we estimate that for each state the uncertainty in the value of $\widetilde{\Delta}_q$ behaves as $\sigma_{\widetilde{\Delta}_q}\lesssim |q| \sigma_\psi/\ln^2\lambda$. A quick analysis leads us to think that in the region $q\geqslant 1$ some points in the data sets might be close to this boundary, and hence it is plausible that their error is slightly underestimated. 

The data set with the highest precision occurs for $\widetilde{\alpha}_0$ ($0.04\%$). From the MFSS analysis of $\widetilde{\alpha}_0$
we find
\begin{equation}
W_c=18.824\; (18.822,18.826)
\end{equation}
and
\begin{equation}
\nu=1.446\; (1.440,1.452)
\end{equation}
where the error limits correspond to $95\%$ confidence intervals. Our estimate for the critical exponent is in perfect agreement with recent results from transfer matrix calculations on the same random phase model.\cite{Slevin2016} 
The value of $W_c$ also agrees reasonably well. 
The study of Ref.~\onlinecite{Ujfalusi2015a} 
using MFSS on a related Hamiltonian belonging to the unitary symmetry class also reports a value of $\nu$ in accordance with our estimate.

\subsection{The multifractal spectrum}

\begin{figure*}
 \centering
 \includegraphics[width=.48\textwidth]{\figdir/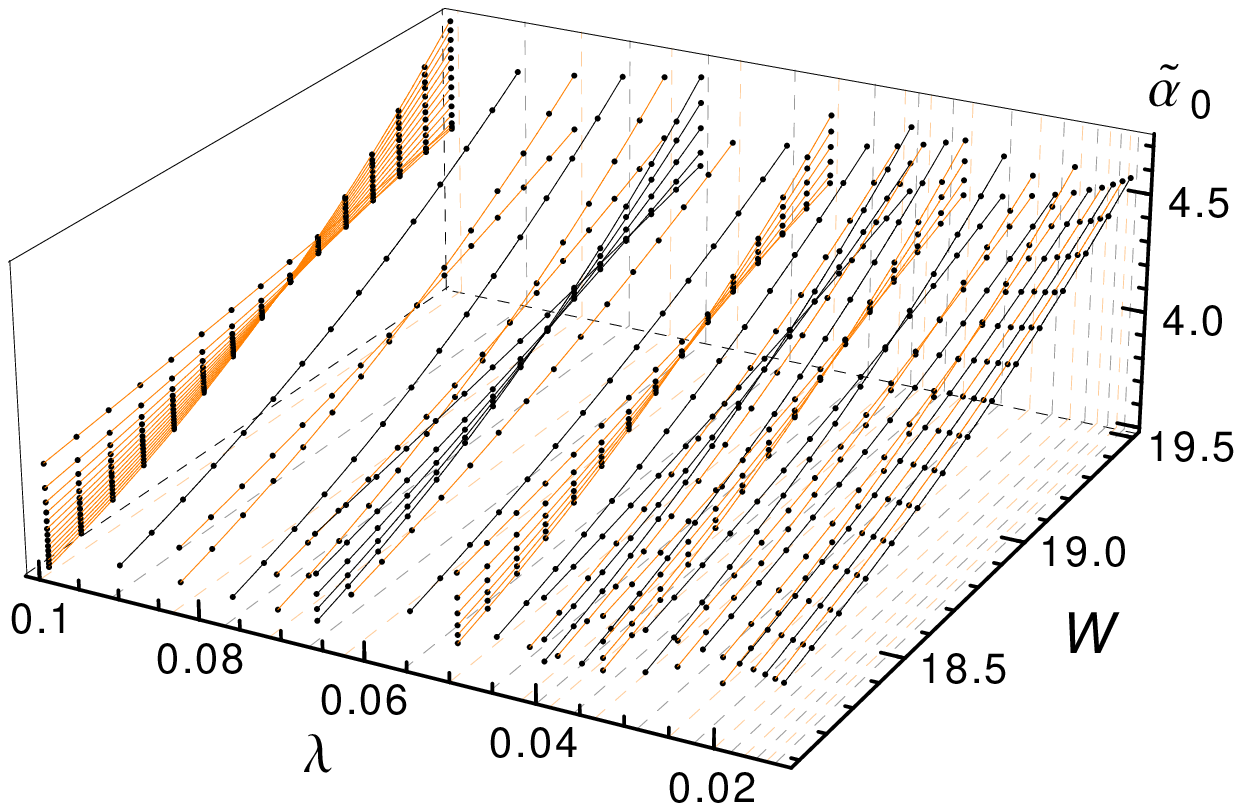}\qquad
 \includegraphics[width=.45\textwidth]{\figdir/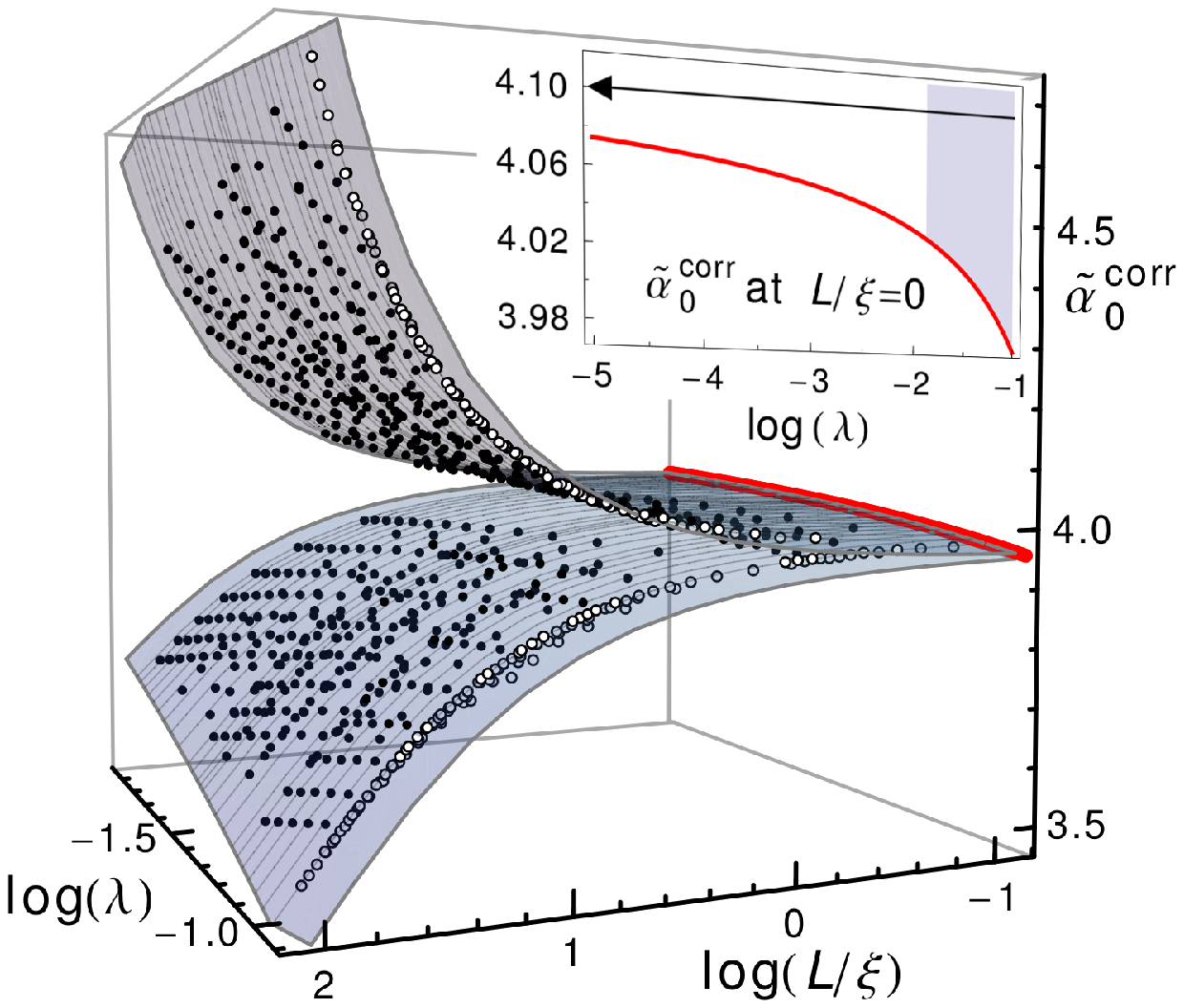}
 \caption{MFSS of $\widetilde{\alpha}_0$. (Left) GMFE (\textbullet) as functions of disorder $W$ for different $\lambda=\ell/L$. The solid lines are cross-sections at fixed $\lambda$ of the best fit, plotted for different $L$. Note that all points are fitted simultaneously. Alternating colors have been used for better visualization.
 (Right) GMFE with irrelevant contribution subtracted (\textbullet, $\circ$) and the scaling surfaces
 (symbol $\circ$ highlights the maximum value of $\lambda$).
 The inset is the scaling function at the critical point, highlighted also in the right face of the main plot.
 The arrow indicates the multifractal exponent given by the extrapolation $\lambda \rightarrow 0$.
 The shaded regions indicate the range of $\lambda$ accessed in our simulations.
 }
 \label{fig:3Dfss}
\end{figure*}

In Fig.~\ref{fig:3Dfss} we show the best fit and the corresponding scaling surface for $\widetilde{\alpha}_0$ in terms of the variables 
$L/\xi$ and $\lambda$. The scale invariant multifractal exponent $\alpha_0$ corresponds to the asymptotic value at the critical point as $\lambda\rightarrow 0$.
This is highlighted in the inset of Fig.~\ref{fig:3Dfss}, where the behavior of the scaling function at criticality ---when the sheets of extended and localized phases meet--- is shown versus $\log(\lambda)$.

\begin{figure}[tb]
 \includegraphics[width=.9\columnwidth]{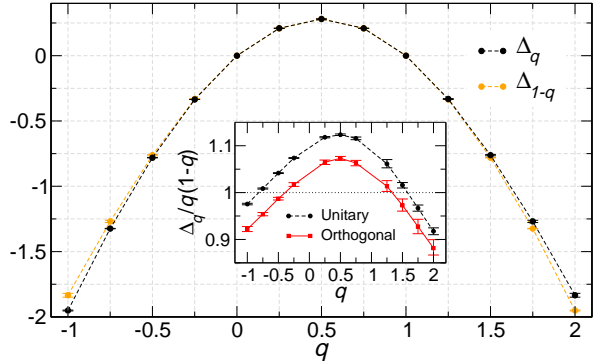}
 \caption{Multifractal exponents $\Delta_q$ obtained from MFSS. The numerical values are listed in Table \ref{tab:results3D}. Error bars denote 95\% confidence intervals. Note that $\Delta_0=\Delta_1=0$ by definition. The inset shows a comparison of the reduced multifractal spectrum $\Delta_q/q(1-q)$ between the unitary and the orthogonal symmetry classes (data from Ref.~\onlinecite{Rodriguez2011}).}
 \label{fig:MFspectrum}
\end{figure}

The whole spectrum of anomalous multifractal exponents, $\Delta_q$, obtained from MFSS is shown in Fig.~\ref{fig:MFspectrum}. As for the orthogonal symmetry 
class,\cite{Vasquez2008,Rodriguez2008,Rodriguez2009,Rodriguez2011} and also in agreement with the results reported in Ref.~\onlinecite{Ujfalusi2015a}, the multifractal spectrum in the unitary symmetry class shows a clear tendency to obey the symmetry relation \eqref{eq:Deltasym}. Nevertheless, within the achievable numerical accuracy, slight deviations from the symmetry become visible as $|q|$ grows. We emphasize that the value of the multifractal exponents is very sensitive to the estimated position of the critical point $W_c$,  which in turn depends strongly on the estimation of $y$. Indeed, obtaining a reliable estimate of $W_c$ is more difficult than extracting the value of the critical exponent $\nu$. Additionally, the extrapolation $\lambda\rightarrow 0$ is intrinsically limited by the range of system sizes available. Therefore, we tend to think that the observed deviations are not a genuine violation of relation \eqref{eq:Deltasym}. Nevertheless, the exact $q$-range for the validity of the symmetry relation would be dependent on the existence (or absence) of termination points in the ensemble averaged multifractal spectrum (see Ref.~\onlinecite{Evers2008}); an issue which is not yet resolved. 

In the inset of Fig.~\ref{fig:MFspectrum}, we show the reduced anomalous exponents $\Delta_q/q(1-q)$ for the unitary and the orthogonal symmetry classes. Both multifractal spectra are remarkably similar (the relative distance between $\Delta_q^\text{unitary}$ and $\Delta_q^\text{orthogonal}$ is around $4$-$5\%$ for $q\in[-1,2]$), and the dominant difference is a shift in the reduced anomalous exponents, in agreement with the observations of Ref.~\onlinecite{Ujfalusi2015a}. As for the orthogonal symmetry class,\cite{Rodriguez2011} the multifractal spectrum exhibits a clear deviation from parabolicity, which requires $\Delta_q\propto q(1-q)$. 
For a comparison of the multifractal spectra against the existing analytical results in $d=2+\epsilon$ dimensions, \cite{Evers2008} we refer the reader to Ref.~\onlinecite{Ujfalusi2015a}, where a thorough analysis is presented.
\section{Conclusions}
\label{sec:Conclusions}
We have presented a detailed numerical analysis of the persistence of multifractal fluctuations in wavefunctions around the 3D Anderson transition in the unitary symmetry class, for a Hamiltonian which describes an electron in a disordered cubic lattice in the presence of a random magnetic flux. The existence of multifractality and the presence of the transition are best observed by analyzing the behavior of the PDF of wavefunction intensities in the vicinity of the critical point. We emphasize that the PDF analysis provides a most convenient way to unambiguously assess the existence of a disorder induced metal-insulator transition.

The use of very large system sizes up to $L^3=150^3$ and averaging over more than 4 million wavefunctions from uncorrelated disorder realizations has led to a very precise estimation of the position of the critical point at energy $E=0$, $W_c=18.824(18.822,18.826)$, and of the localization length critical exponent, $\nu=1.446(1.440,1.452)$.
This latter value is in agreement with previous studies of the same model using the transfer matrix technique \cite{Slevin2016,Kawarabayashi1998a,Kawarabayashi1998,Chalker1995,Henneke1994} 
and level statistics,\cite{Garcia-Garcia2007} and also with results for a system in a uniform magnetic field using transfer matrix,\cite{Slevin1997}
or (as in this work) a generalized multifractal analysis.\cite{Ujfalusi2015a,Lindinger2015T}
We note the small relative difference ($\sim 4\%$) between our estimate for $\nu$ and the analytical result of Ref.~\onlinecite{Garcia-Garcia2008} ($\nu=3/2$), which, at the level of approximation there considered, might be relevant for the unitary symmetry class. \cite{Garcia-GarciaPrivate}

The values for the multifractal exponents reported in Refs.~\onlinecite{Ujfalusi2015a,Lindinger2015T} are compatible with our findings here, and thus this analysis helps to confirm the universality of the properties of the critical point in this symmetry class. 

\begin{acknowledgments}%
We thank Andreas Buchleitner for a careful reading of the manuscript, and 
Antonio Garc\'ia-Garc\'ia for helpful discussions.
The authors acknowledge support by the state of Baden-W{\"u}rttemberg through bwHPC.
\end{acknowledgments}
\bibliographystyle{prsty}

\begin{thebibliography}{10}
\bibitem{PDF}
It must be emphasized that the PDF exposes the presence of the transition also
  under scaling with $L$ at fixed $\ell $. In this case the whole PDF is not
  scale invariant at the critical point, but only the position of its maximum
  is.
\end{thebibliography}

\end{document}